# Spectral Sensitivity of Graphene/Silicon Heterojunction Photodetectors

Sarah Riazimehr[1], Andreas Bablich[1], Daniel Schneider[1], Satender Kataria[1], Vikram Passi[1],

Chanyoung Yim[2], Georg S. Duesberg[2] and Max C. Lemme[1]

[1]University of Siegen, School of Science and Technology, Hölderlinstrasse 3, 57076 Siegen, Germany

[2]Trinity College Dublin, School of Chemistry and Centre for Research on Adaptive Nanostructures and Nanodevices (CRANN) and Advanced Materials and BioEngineering Research (AMBER) Centre, Ireland

max.lemme@uni-siegen.de

*Abstract*—We have studied the optical properties of two-dimensional (2D) Schottky photodiode heterojunctions made of chemical vapor deposited (CVD) graphene on n- and p-type Silicon (Si) substrates. Much better rectification behavior is observed from the diodes fabricated on n- Si substrates in comparison with the devices on p-Si substrates in dark condition. Also, graphene – n-Si photodiodes show a considerable responsivity of 270 mAW$^{-1}$ within the silicon spectral range in DC reverse bias condition. The present results are furthermore compared with that of a molybdenum disulfide (MoS$_2$) – p-type silicon photodiodes.

*Keywords — graphene; molybdenum disulfide; Schottky barrier diode; photodiode; sensitivity; responsivity; spectral response.*

I. INTRODUCTION

A broad spectral range of detection is important for many photonic applications such as imaging, sensing, night-vision, motion detection, communication and spectroscopy [1], [2]. Recently, graphene has attracted significant scientific attention for ultra-broadband optoelectronic applications since it offers numerous advantages compared to the other commercially available photosensitive materials. First of all, graphene is able to absorb 2.3 % of the incident light despite being only one atom (0.34 nm) thick [3]. Such an absorption is observed in 20 nm thick silicon (Si) and 5 nm thick Gallium-Arsenide (GaAs), the two most commonly used materials in optoelectronic applications [4]. Regardless of its impressive light absorption, single layer graphene is still far below the requirements of optical applications. Furthermore, graphene exhibits a very broad spectral range of detection from ultraviolet to terahertz, and quasi wavelength independent absorption, which is a consequence of its gapless and linear dispersion relation [3], [5]. In addition, graphene shows an extremely high charge carrier mobility which enables ultrafast photodetection [6],[7],[8]. Moreover, compatibility of graphene as a material with the well-established Si process line makes it a promising candidate for large-scale and cost-effective technological applications [9], [10]. Over the past few years, there have been numerous studies on both, photovoltaic and photo-thermoelectric effects in graphene devices [6], [7], [11]–[17]. Besides, complex architectures including asymmetric metal electrodes [7], plasmonic nanostructure [18], [19] and microcavities [20], [21] have been used in the earlier studies in order to enhance the device photocurrent. Nevertheless, responsivity above 21 mAW$^{-1}$ could not be achieved. The low responsivities are attributed to the weak total optical absorption of a single layer of graphene. It should be noted that most of the mentioned studies were focused on mechanically exfoliated graphene devices, which are not suitable for large-scale device integration.

Molybdenum-disulfide ($MoS_2$), a layered transition metal dichalcogenide (TMD), is an important member in the family of two-dimensional (2D) materials. In contrast to graphene, $MoS_2$ is a semiconductor and its bandgap varies depending on the number of the layers. Therefore, $MoS_2$ has the possibility to detect light at different wavelengths. Monolayer $MoS_2$ has a direct bandgap of ~1.8 eV, while bulk $MoS_2$ has an indirect bandgap of ~1.3 eV [22]. In addition, $MoS_2$ has high transparency and mechanical flexibility and it is easy to process [23]. All of these exceptional properties make $MoS_2$ another promising material for electronic and optoelectronic applications.

In this work, we report on CVD graphene-based Schottky barrier photodiodes of simple architecture, which are also scalable, reproducible and of low-cost. To fabricate the diodes, we have adopted a similar process as in [24], [25], where large area CVD-grown graphene films were transferred onto pre-patterned n- and p-type Si-substrates. Afterwards, we performed electrical and optical characterizations on the photodiodes. The optical data obtained from the measurements are further compared to that of $MoS_2$ – p-type silicon photodiodes, which have been presented in detail in our previous work [26].

## II. EXPERIMENT

### A. Device Fabrication

Lightly doped p- and n-type Si <100> wafers with a thermally grown silicon dioxide ($SiO_2$) layer of 85 nm were used as starting substrates. The p-Si wafer was boron doped with a doping concentration of $3\times10^{15}$ cm$^{-3}$ and the n-Si wafer was phosphorus doped with a doping concentration of $2\times10^{15}$ cm$^{-3}$. These wafers were diced into 13×13 mm$^2$ square samples for device fabrication. Eight photodiodes were fabricated on each sample. After a first standard UV-photolithography step, the oxide was partially etched with buffered oxide etchant (BOE) for 90 seconds to expose the silicon. The contact metal electrodes were defined by a second photolithography step followed by sputtering of 20 nm chromium (Cr) and 80 nm gold (Au) and lift-off process. The electrodes were deposited immediately after the native oxide removal in order to form good ohmic contacts. Large-area graphene was grown on a copper foil in a NanoCVD (Moorfield, UK) rapid thermal processing tool by the CVD method. To transfer graphene films onto pre-patterned substrates, ~1 cm$^2$ pieces of graphene coated Cu foil were spin-coated with polycarbonate (PC) (Bisphenol A) and baked on a hot plate at 85 °C for 5 minutes. Electrochemical delamination (bubble transfer method) was performed to remove the polymer-supported graphene films from the copper surface (see details in [27]). Prior to graphene transfer, the native silicon oxide on silicon substrates was removed by BOE, ensuring good electrical contact between graphene and Si substrate. Then, the graphene films were transferred on top of the pre-patterned substrates, ensuring that they cover the electrodes. Afterwards, the devices were thoroughly immersed into chloroform overnight, followed by cleaning them with acetone, isopropanol and DI water and drying. Finally, a last photolithography step was performed followed by oxygen plasma etching of graphene to define graphene junction areas ranging from 0.64 mm$^2$ to 1.6 mm$^2$. The different fabrication steps are shown in Fig. 1.

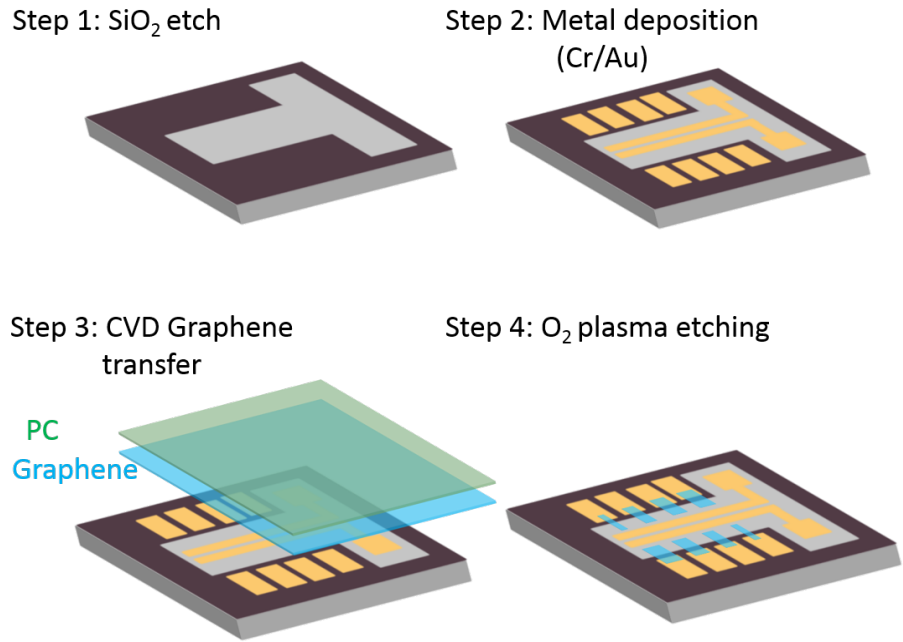

Figure 1: Fabrication process of graphene photodiodes. Step 1: BOE etching SiO$_2$. Step 2: Sputtering of electrodes after native oxide etching. Step 3: Polymer-supported graphene transfer. Step 4: Polymer removing in chloroform followed by photolithography and graphene oxygen plasma etching.

## B. Electrical Characterization

Electrical measurements on the diodes were performed on a Karl Süss probe station connected to a Keithley semiconductor analyzer (SCS4200) under ambient conditions. The voltage for all devices was swept from 0 to +2 V for forward and from 0 to -2 V for reverse biasing. A white light source (50 W halogen lamp) with a dimmer to control the light intensity was used to measure the photoconductivity of the diodes. The variable incident light intensity has been modulated between 0 and 3 mWcm$^{-2}$. A CA 2 laboratory thermopile was used to measure the intensity of the light source.

*C. Optical Characterization*

The spectral response (SR) of the photodetectors was measured by comparing it to calibrated reference detectors using a lock-in technique. A tungsten-halogen and deuterium-arc lamp (wavelength ranging between 200 nm and 2200 nm) were used to generate light. Specific wavelengths were selected by a monochromator. The light power density varied from 1 µWcm$^{-2}$ at a wavelength of 200 nm up to 55 µWcm$^{-2}$ at a wavelength of 1150 nm. A chopper with a frequency of 17 Hz was used to modulate the intensity of the light beam. Calibrated Si and Indium-Gallium Arsenide (InGaAs) diodes were measured as reference. The photocurrents were recorded through pre-amplifiers (FEMTO) and lock-in amplifiers at 17 Hz chopper frequency for detection of ultra-low currents down to 10 pA. The measurement principle allows to establish a wavelength dependent correction factor for the responsivity calculation which takes into account variations of the preamplifiers, varying photo flux densities caused by the monochromator as well as the area difference between the reference detector and the sample.

## III. RESULTS AND DISCUSSION

In order to fabricate graphene photodiodes, graphene films of ~ 1 cm$^2$ area were transferred on to pre-patterned p- and n-Si substrates, as described in the experimental part. Fig. 2 shows a schematic and a scanning electron micrograph of a graphene – Si device. One end of the structured graphene film is in contact with an Au pad on SiO$_2$ forming an ohmic contact. The other end is in contact only with the Si substrate where it forms a p-n junction. The metal pads, SiO$_2$, and graphene are shown in yellow, purple and blue, respectively.

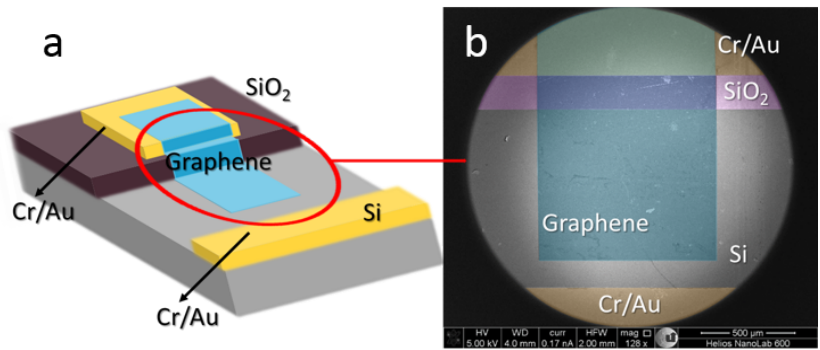

Figure 2: (a) Schematic and (b) scanning electron micrograph of a graphene photodiode. The SEM image is color enhanced to show the exact position of the graphene film, SiO$_2$, and metal electrodes.

The current density-voltage (J-V) characteristics of both, n- and p-Si – graphene devices, are shown in Fig. 3a and b, respectively. The graphs show measurements on four different chips and each color represents one chip; black, red, green and blue for chip number 1, 2, 3 and 4, respectively. For each chip, several diodes have been measured. Although these results represent devices on four different chips, devices of the same type show similar characteristics and data reproducibility. The fabricated devices on n-Si substrates exhibit a clear rectifying behavior while the devices fabricated on p-Si substrates show non-rectifying behavior. This behavior is attributed to ambient p-type doping of the graphene caused by air and humidity molecules, leading to a p-n junction on n-Si and a p-p heterojunction on p-Si [28], [29].

Henceforth, further measurements in this work were performed on the graphene – n-Si diodes. The forward J-V characteristic of the diode can be described by the single-exponential Shockley equation [30] allowing the extraction of the ideality factor and the Schottky barrier height (SBH). The ideality factor, the Schottky barrier height (SBH) and the series resistance of $1.52 \pm 0.1$, $0.65 \pm 0.08$ eV and $7.5 \pm 0.7$ k$\Omega$, respectively, were extracted for the diodes in Fig. 3a. The extraction method is described in detail in [31]. The corresponding energy band diagrams for graphene – n-Si and p-Si diodes at zero bias voltage in the dark is shown in Fig. 3c and 3d, respectively.

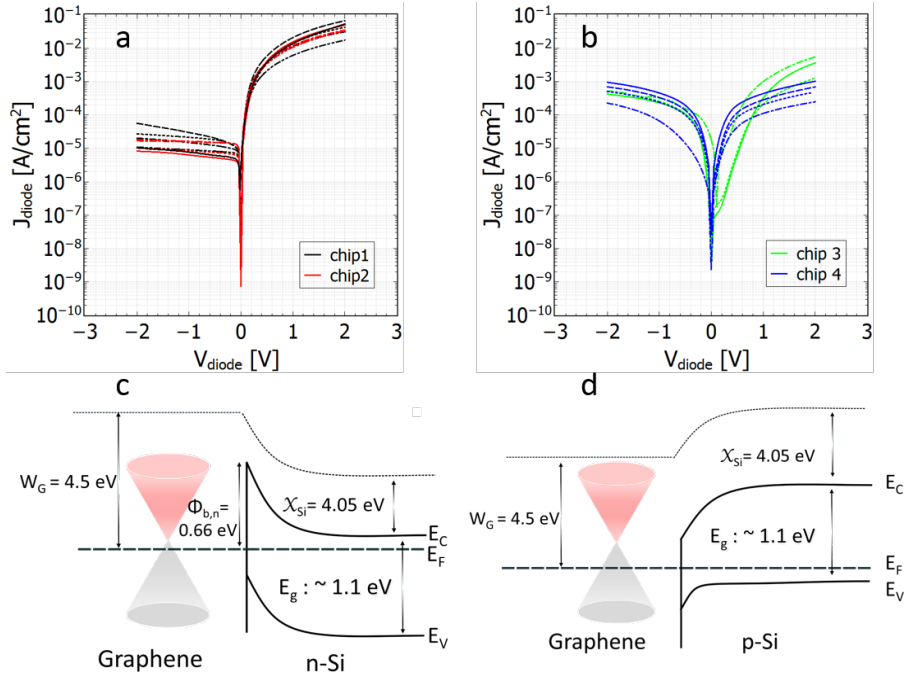

Figure 3: (a) Current density-voltage (J-V) characteristics of the graphene – n-Si photodiodes and (b) graphene – p-Si junctions. Measurements were performed on four different chip and their plots are shown in black, red, green and blue for chip number 1, 2, 3 and 4, respectively. There are eight diodes per chip and each trace belongs to different diode. Schematic band diagram of (c) the graphene – n-Si interface (d) and the graphene – p-Si interface at zero bias voltage. $E_C$, $E_V$, $E_F$, $E_g$, $W_G$, $\chi_{Si}$ and $\Phi_b$ indicate conduction band, valence band, Fermi level, bandgap, graphene work function, Si electron affinity and Schottky barrier height, respectively.

The photoconductivity of the graphene – n-Si diodes has been further investigated under white-light illumination conditions. Fig. 4a and 4b show the J-V plots of one representative diode in the dark and under illumination in the linear and semi-logarithmic scale. The diode is in the off-state under reverse bias in the dark and exhibits low dark current density of 13.4 $\mu Acm^{-2}$, while under illumination it displays a noticeable photocurrent density of 4.5 $mAcm^{-2}$. Thus, about three orders of magnitude change in current density between dark and illuminated conditions in the reverse bias and a minor variation under forward bias is obtained. Furthermore, the device shows a high sensitivity to variation of light intensity, as shown in Fig. 4c. An increase in light intensity from zero to maximum, results in increasing current density from 13.4 $\mu Acm^{-2}$ to 4.5 mAcm$^{-2}$. The linear dynamic range of the photodetector was also measured from dark to full light intensity to be 3.7 mWcm$^{-2}$ at a reverse dc bias of 2 V. As shown in Fig. 4d, the generated photo current density shows a nearly linear increase with the incident light intensity changing from 0 mWcm$^{-2}$ (dark) to 3.7 mWcm$^{-2}$, corresponding to a linear dynamic range of 43 dB. This low linear dynamic range is attributed to a very high dark current of graphene-based photodetectors due to graphene's gapless nature. A cross-section of a graphene – n-Si diode and the corresponding energy band diagram in the reverse bias under illumination is shown in Fig. 5. When the graphene – n-Si substrate junction is illuminated, the incident photons generate electron-hole pairs in the n-Si and in the graphene. By applying a reverse bias, the photo-generated holes in Si are accelerated into graphene, leading to a significant photocurrent [32].

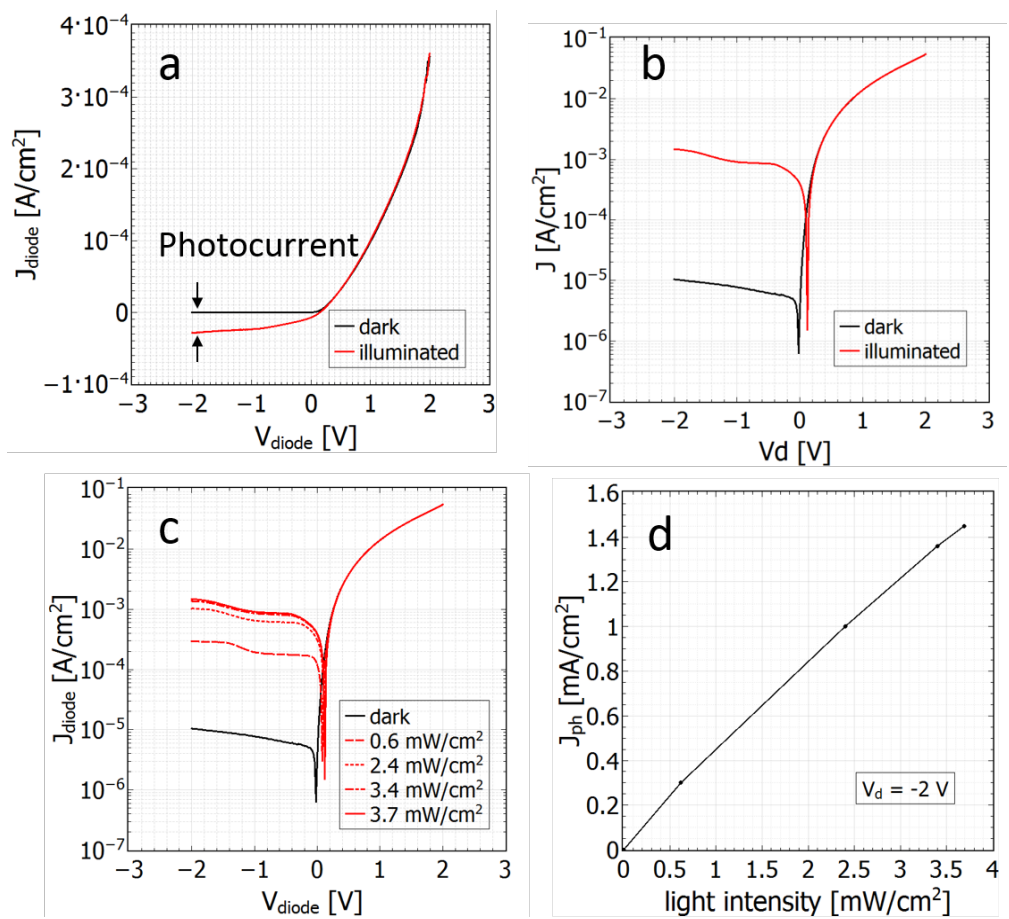

Figure 4: J-V plot of the graphene – n-Si diode on (a) linear and (b) semi-logarithmic scale in the dark and under illumination. (c) J-V plot of the graphene – n-Si diode on the semi-logarithmic scale under various light intensities of 0.6, 2.4, 3.4 and 3.7 mWcm$^{-2}$, compared to the dark state. (d) Photocurrent density of the device with varying light intensity at reverse bias voltage of 2 V.

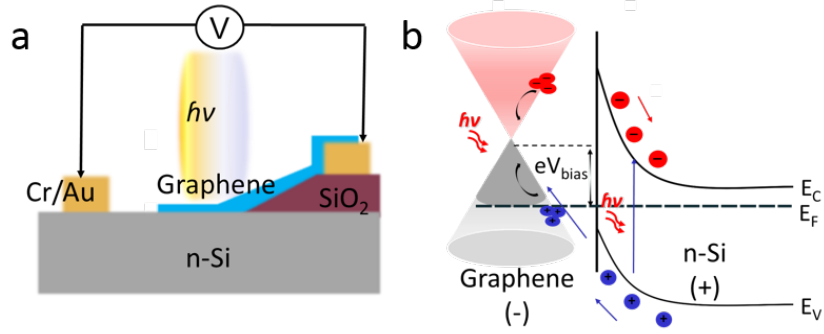

Figure 5: (a) Cross-section of the graphene – n-Si heterojunction diode and (b) its band diagram in reverse biased condition under illumination.

The absolute SR was measured using an automated lock-in technique by a LabVIEW controlled set-up. The SR of seven different graphene – n-Si diodes at applied reverse bias voltages of 2 V are plotted in Fig. 6a. All the diodes exhibit a broad SR ranging from ultraviolet (UV) to near infrared (NIR), showing a nearly identical behavior. From Fig. 6a, the position of the maximum responsivity is extracted to be between 195 mAW$^{-1}$ and 270 mAW$^{-1}$ at approximately eV 1.33 eV (930 nm). The obtained maximum responsivity of 270 mAW$^{-1}$ is slightly less than that reported by An *et al.* which was around 300 mAW$^{-1}$ at 720 nm [32]. This minor difference, which is less than the variation of the responsivity observed in our case, may be attributed to an inhomogeneous coverage of the transferred graphene due to holes and tears resulting from the graphene transfer. In addition, polymer residue in some parts of the graphene films may also contribute to variability. We therefore believe that this observed variation can be reduced by improving the graphene transfer process. Another possible reason for the different performance in terms of responsivity values between the presented data and the data reported in [32] may be due to different geometries of contact electrodes. Larger contact electrodes with a vertical configuration in [32] result in an increased external electrical field. Therefore, more photo-generated electron-hole pairs may be captured before recombining and consequently, the overall responsivity is higher. The main photoexcitation resides in the Si, and therefore one expects to see SR similar behavior as in silicon pn-diodes. We have indeed observed a peak at 930 nm, which is close to the Si absorption peak. We also achieved a broader operational bandwidth (in the Si operating wavelength region) with a single layer graphene – n-Si diode, while in the previous publication [32] similar results require 3 layers of graphene. This may be due to the different p doping levels in the graphene samples as a result of different laboratory conditions. Graphene p doping is expected to increase its sheet conductance, which has been previously used to enhance the performance of graphene – Si Solar cells [33]. SR measurements over a broader spectrum at various applied reverse bias voltages were done on the same graphene – n-Si diode, for which the data is plotted in Fig. 4. The SR measurement plot shows an increase of the SR with the applied reverse bias due to the increment of the electrical field (Fig. 6b). In this wide range from 360 nm to 2000 nm, the responsivity varies by several orders of magnitude, between 0.18 mAW$^{-1}$ and 230 mAW$^{-1}$ at a reverse dc bias of 2 V. The observed peak at approximately 1.33 eV (930 nm) is due to absorption in the underlying n-type silicon. In fact, the diodes exhibit a SR similar to silicon p-n diodes, even though one doped region has been replaced by monolayer graphene. Its responsivity is about 50% of that of the commercial calibrated reference diode. Below the cutoff frequency of Si, i.e. for energies below the silicon bandgap, where there is no contribution from the underlying n-type Si, we observed a low and flat optical response over a broad spectrum as a result of absorption in the graphene layer. The responsivity drops to less than 0.2 mAW$^{-1}$. In graphene, electron-electron scattering can lead to photo-assisted carrier multiplication, increasing the number of electrons that might contribute to a photocurrent before recombination occurs [34]. Nevertheless, the absolute number of optically excited carriers in the graphene that contribute to the photocurrent is much less than in the Si, and therefore the low responsivity in this range is in line with light absorption of up to 2.3 % for single layer graphene [3]. These results demonstrate that graphene not only acts as a carrier collector, as reported previously, but also as an absorber.

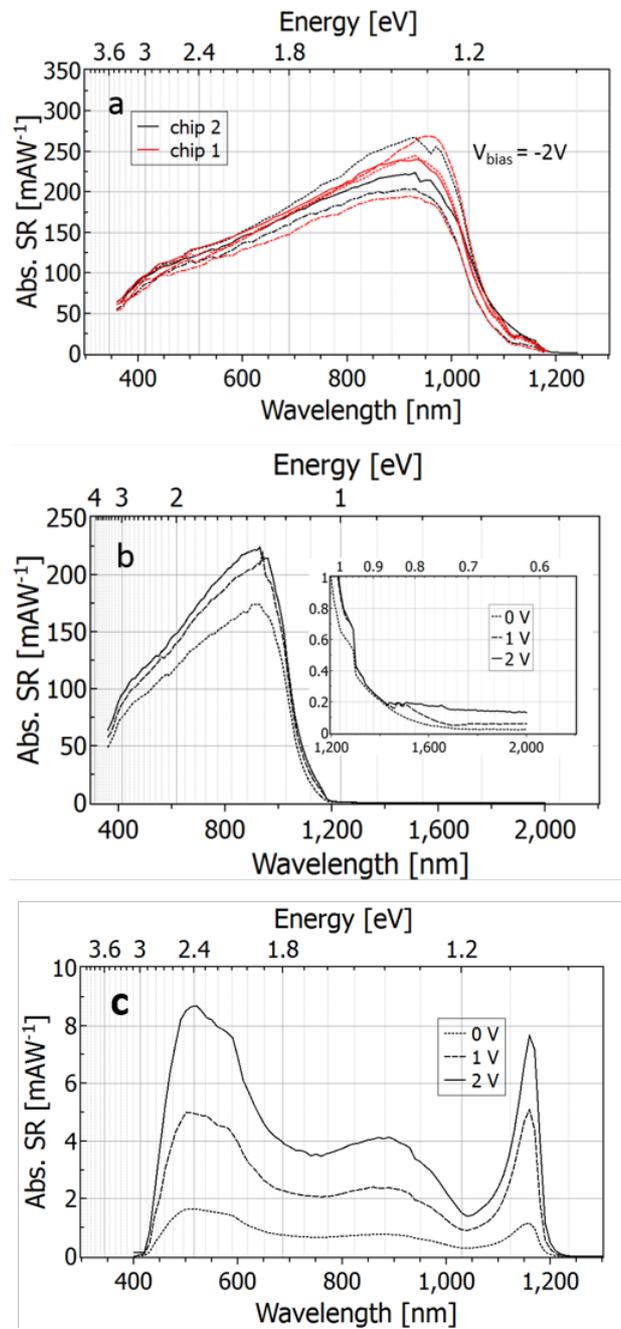

Figure 6: Absolute spectral response (Abs. SR) vs. wavelength (lower x-axis) and energy (upper x-axis) of (a) different graphene - n-Si photodiodes at a reverse bias of 2 V. Measurements were done on two different chips and their plots are shown in black and red for chip number 1 and 2, respectively, (b) a graphene - n-Si photodiode for wavelengths ranging from 360 nm (3.44 eV) to 1200 nm (1.03 eV), the inset shows zoom-in from 1200 nm (1.03 eV) to 2100 nm (5.9 eV) and (c) an $MoS_2$ - p-Si photodiode at zero bias and reverse biases of 1 V and 2 V.

Finally, the SR of the graphene – n-Si diodes is compared with previously reported CVD $MoS_2$ – p-Si photodiodes [26]. In contrast to the gapless graphene, molybdenum disulfide ($MoS_2$) is an n-type two-dimensional semiconductor material. An $MoS_2$ – p-Si diode therefore displays a different spectral response, which is limited by the bandgap of $MoS_2$. The SR plot of the multilayer $MoS_2$ – p-Si diode is shown in Fig. 6c (adapted from [26]). Responsivities between 1.4 $mAW^{-1}$ and 8.6 $mAW^{-1}$ at a reverse DC bias of 2 V are achieved in the range from 400 nm to 1200 nm. The absorption peak at approximately 1.1 eV (1127 nm) is related to the underlying p-type Si substrate. Three additional absorption peaks reveal the band structure of multilayer $MoS_2$ with an indirect band transition of ~1.43 eV (867 nm) and two direct band transitions located at ~2.15 eV (576 nm) and ~2.48 eV (500 nm). We observe a blue-shift of 0.13 eV for the indirect transition and 0.4 eV for the two direct band transitions compared to literature data (i.e. 1.3 eV for the indirect band transition; 1.8 eV and 2.0 eV for the direct band transitions [22], [35], [36]) on exfoliated $MoS_2$ devices. The observed blue shift is attributed to a lattice compression of CVD $MoS_2$ based on experimental and theoretical evidence as described in detail in [26]. A reported maximum responsivity of 8.6 $mAW^{-1}$ in [26] is due to the absorption in multilayer $MoS_2$. This value is less than the overall responsivity value obtained in the graphene – n-Si diodes (mainly resulting from Si), but it is more than the absorption exhibited by monolayer graphene.

IV. CONCLUSIONS

Graphene based Schottky barrier photodiodes were fabricated using a simple, scalable and reproducible technology. We observed that diodes fabricated on an n-Si substrate exhibit a rectifying behavior, whereas the devices fabricated on a p-type Si substrate act as photoresistors. This behavior can be attributed to p-type graphene-doping due to its exposure to the ambient atmosphere leading to a higher Schottky barrier to the n-type silicon. Graphene – n-Si diodes exhibit a broad SR with responsivity up to 270 $mAW^{-1}$. Nevertheless, the main photoexcitation takes place in the n-type Si substrate due to low optical absorption by single layer graphene. Unlike in graphene-based diodes, the SR is limited to the band structure of $MoS_2$ in the $MoS_2$- based photodiodes. This study promotes SR measurements as an excellent tool to probe the electronic properties of novel 2D-materials.

ACKNOWLEDGMENTS

Support from the European Commission through an ERC starting grant (InteGraDe, 307311) as well as the German Research Foundation (DFG, LE 2440/1-1 and GRK 1564) is gratefully acknowledged. The authors would like to thank Dr. Heiko Schaefer-Eberwein (Univ. of Siegen) for SEM studies. GSD and CY acknowledge SFI under Contract No. 12/RC/2278 and PI_10/IN.1/I3030 and the European Union Seventh Framework Programme under grant agreement No. 604391 Graphene Flagship.